\documentclass[10pt,oneside]{Lowx2013}
\usepackage{epsf,amsmath}
\usepackage{graphicx}
\usepackage{caption}
\usepackage{subfigure}
\usepackage{epsfig}

\begin{document}
\bibliographystyle{unsrt}

\title{Multiparton Interactions and Double Parton Scatterings in CMS}

\author{\underline{Paolo Gunnellini}, on behalf of the CMS Collaboration
 \\ \\
DESY, Notkestrasse 85, 22761 Hamburg, Germany\\
e-mail: paolo.gunnellini@desy.de\\
{\bf Contribution given at the low-x workshop,} \\
{\bf May 30 - June 4 2013, Rehovot and Eilat, Israel}}

\maketitle

\begin{abstract}
\noindent  Multiparton interactions are introduced in order to explain a wide range of phenomena in p-p collisions. We present the most recent CMS measurements sensitive to multiparton interactions and hard double parton scattering at 7 TeV. In particular, the W+dijet channel and a scenario with four light jets in the final state are described.
\end{abstract}



\markboth{\large \sl \hspace*{0.25cm}\underline{C. Royon} 
\hspace*{0.25cm} } {\large \sl \hspace*{0.25cm} Low-x 2013 Proceedings}

\section{Introduction}
The basis for understanding hadronic collisions at high energy is provided by the QCD improved parton model. Due to the high gluon density in each colliding hadron and the increasing parton cross section, it is possible to have multiple parton interactions (MPI)(\cite{Bartalini:2011jp}) where two or more distinct hard parton interactions occur simultaneously in a single collision. With increasing collision energy, in particular, the contribution of these additional scatterings become more and more relevant because partons with successively lower longitudinal momentum fraction x become active and can be probed. An interaction where each proton has two active partons giving rise to two different hard subprocesses is usually referred as a Double Parton Scattering (DPS). Thanks to the high collision energy and luminosity, the Large Hadron Collider (LHC) provides a valuable opportunity to observe direct DPS evidence in several physics channels. The Compact Muon Solenoid (CMS \cite{Chatrchyan:2008aa}) experiment has carried out several measurements in different physics channels at a center of mass energy of 7 TeV, in order to study the activity and the contribution of both soft and hard MPI.

\section{Soft MPI measurements}
A complete description of the hadronic activity in high energy collisions at the LHC requires the understanding of the Underlying Event (UE), as it constitutes an unavoidable background to most observables. From an experimental point of view, the UE gathers all the activity accompanying the actual hard scattering. In particular, MPI constitutes the main component of the UE, along with the additional parton radiation (PS) and the interactions between constituents of beam remnants, left behind after the scattered partons have been pulled out. Since it is impossible to separate the UE from the hard scattering process on an event-by-event basis, the topological structure of the measured final state has to be investigated. Typically, studies of the UE rely on measurements of the properties of charged particle production: by observing them, one may investigate the region at very low $p_T$, crucial for exploring soft and semi-hard physics. At higher values of $p_T$, it is possible to directly observe hard MPI in the form of DPS, described in section 3.\\
The measurement of the UE is usually performed by looking at charged tracks, measured with the inner tracking system, in different regions of the $\eta$-$\phi$ plane. These regions are defined according to the direction of the physical object associated with the hard scattering: for example, this might be the leading track when studying QCD interactions or a di-lepton pair when looking at Drell-Yan events. These objects are needed also to set the scale of the hard scattering and study the dependence of the UE. The azimuthal angular difference between charged tracks and the hard scale object, $|\Delta\phi|$ = $|\phi-\phi_{hard\mbox{ }object}|$, is used to define the following three azimuthal regions:
\begin{itemize}
\item $|\Delta\phi|$ $<$ 60$^\circ$, ``toward region'';
\item 60$^\circ$ $<$ $|\Delta\phi|$ $<$ 120$^\circ$, ``transverse region'';
\item $|\Delta\phi|$ $>$ 120$^\circ$, ``away region'';
\end{itemize}


The features of these regions are extensively described in \cite{Bartalini:2011jp}. In the transverse region, the UE is expected to be more important since there, the contribution from the hard scattering is expected to be minimal. The activity in the transverse region has been measured in CMS in two different types of events: QCD processes (\cite{CMS:2012zxa}) where the direction of the leading track has been taken as reference for the hard scattering and Drell-Yan processes (\cite{Chatrchyan:2012tb}) where the di-muon pair, selected in the Z mass range between 81 and 101 GeV/$c^2$, defines the scale of the interaction. Charged tracks in $|\eta|$ $<$ 0.8 for the QCD events (for consistency with the ALICE measurement \cite{ALICE:2011ac}) and in $|\eta|$ $<$ 2.0 for the Drell-Yan UE with $p_T$ $>$ 0.5 GeV/c have been measured in the transverse region as a function of the scale of the hard process:
\begin{itemize}
\item average number of primary charged particles per unit of pseudorapidity and of azimuthal angle; 
\item average of scalar sum of the transverse momenta of primary charged particles per unit of pseudorapidity and azimuth;
\item ratio of the previous two quantities.
\end{itemize}        

\begin{figure}[htbp]
\begin{center}
\subfigure[]{\includegraphics[width=3.34cm]{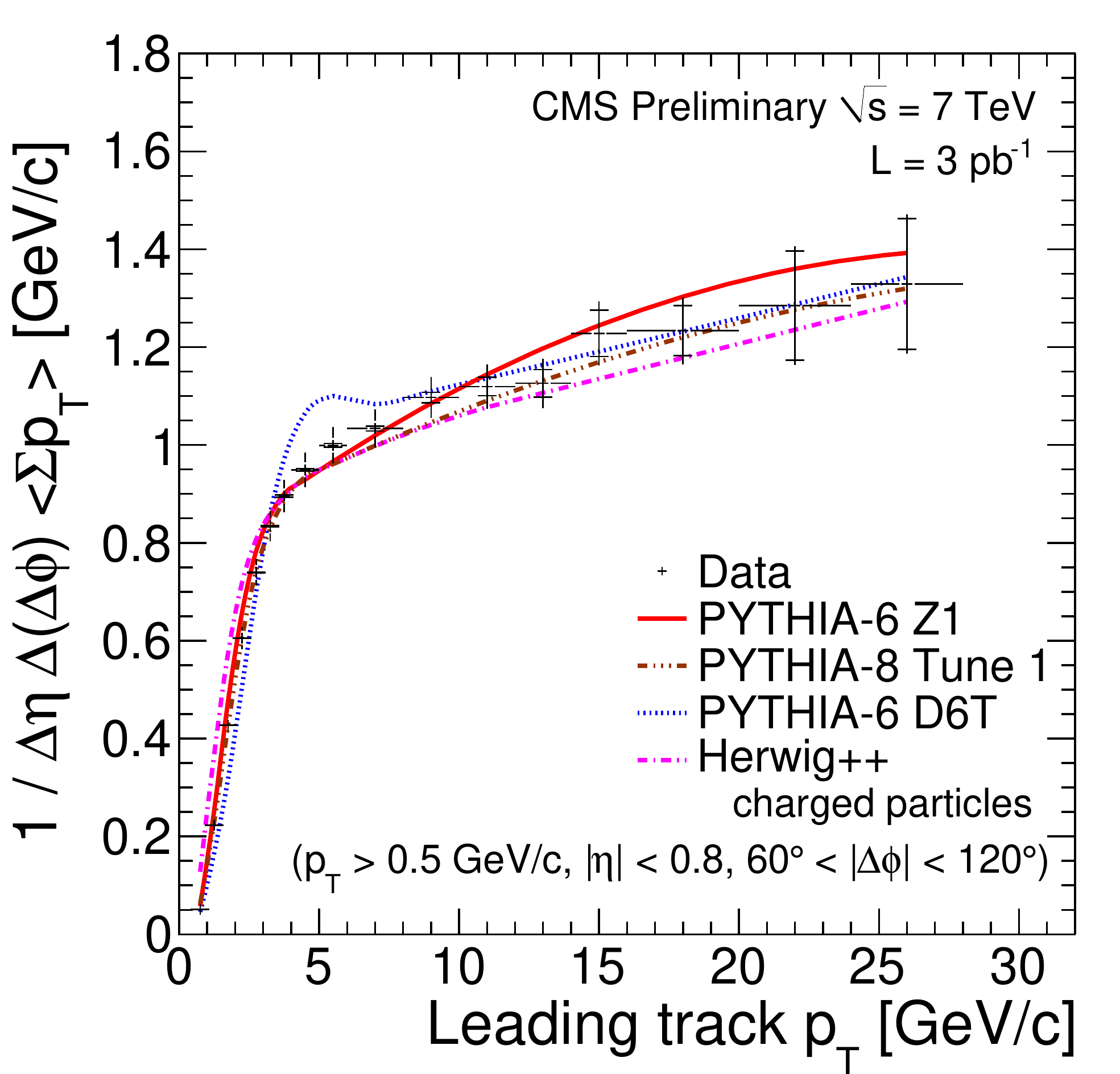}}
\subfigure[]{\includegraphics[width=3.34cm]{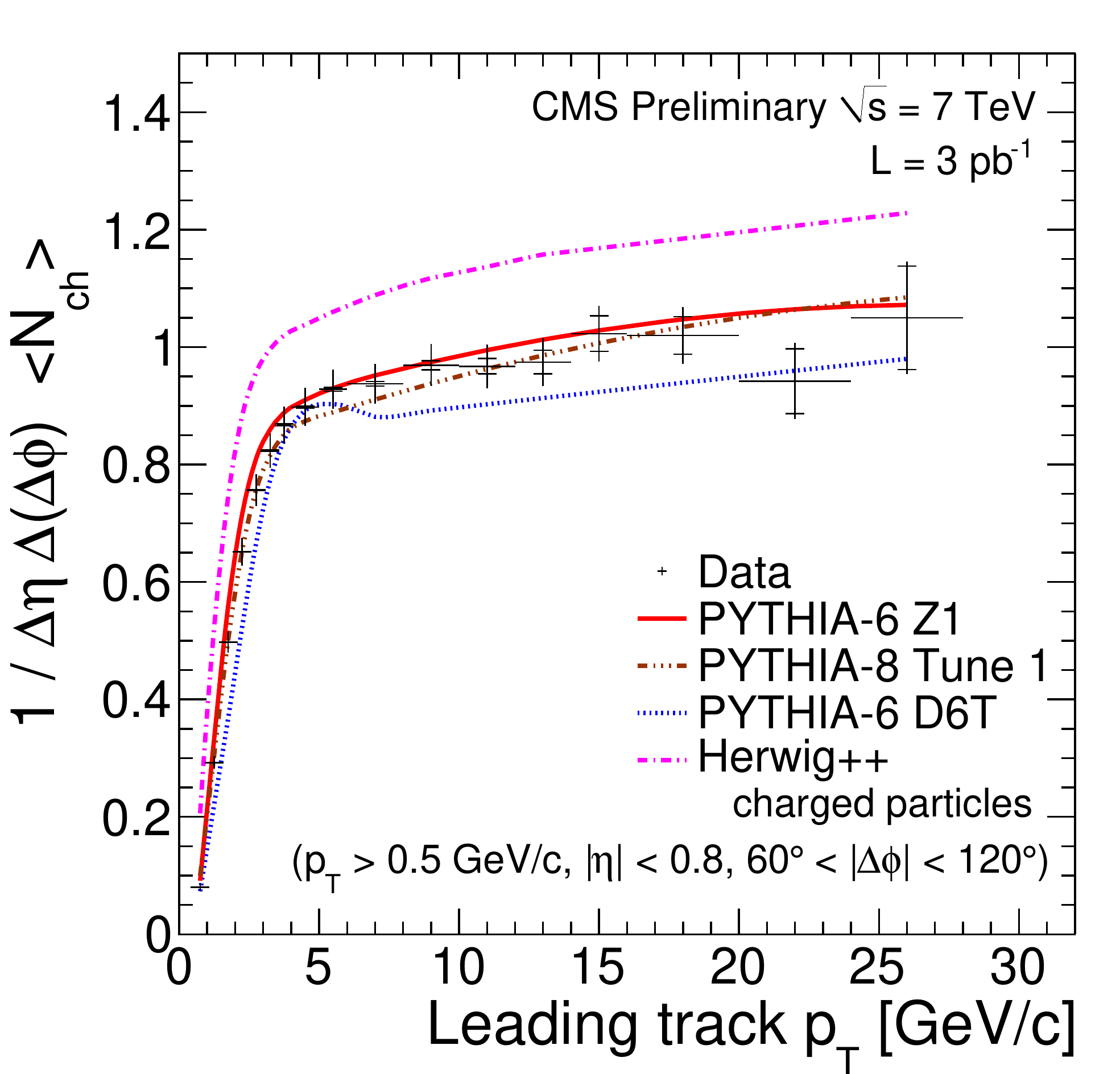}}
\subfigure[]{\includegraphics[width=3.25cm]{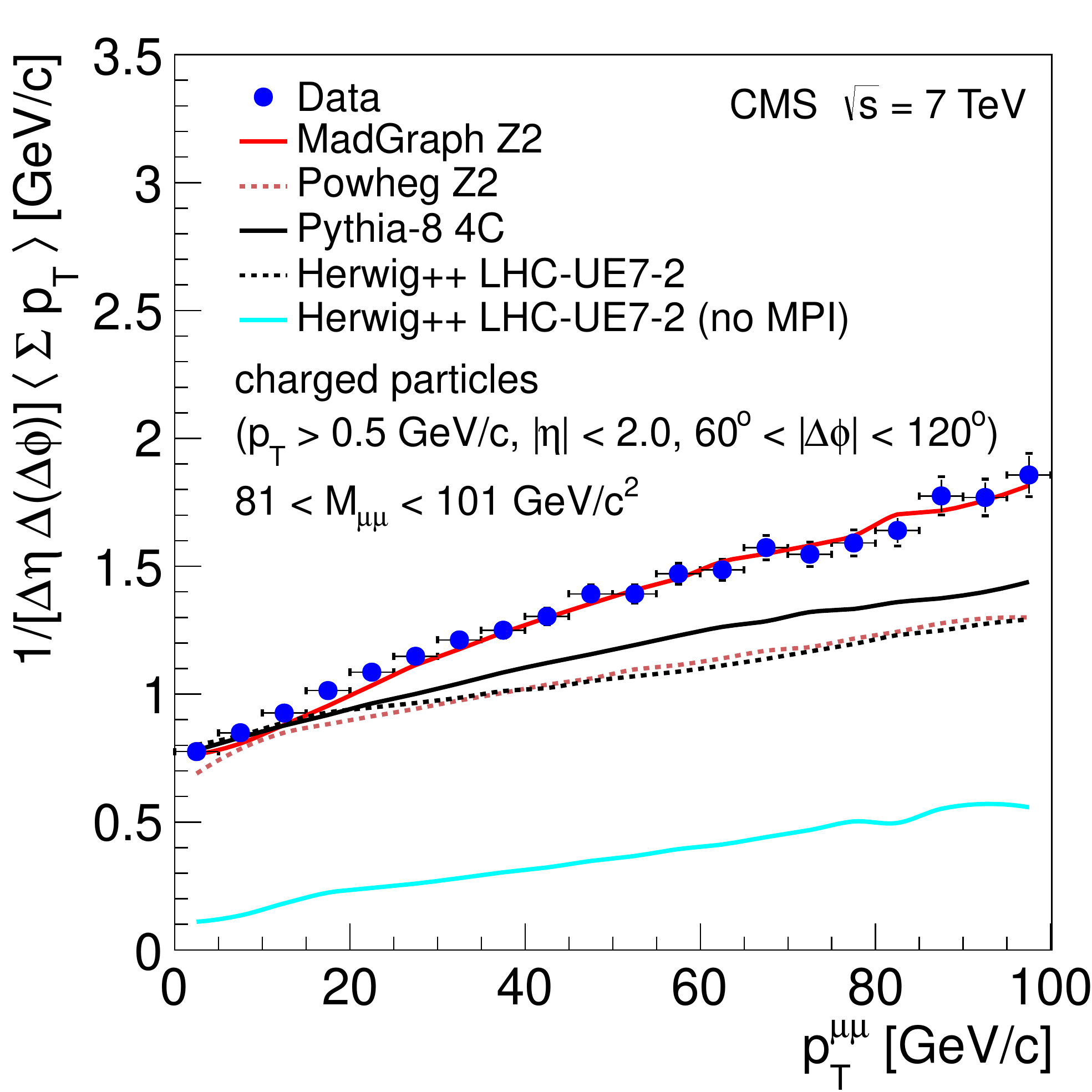}}
\subfigure[]{\includegraphics[width=3.25cm]{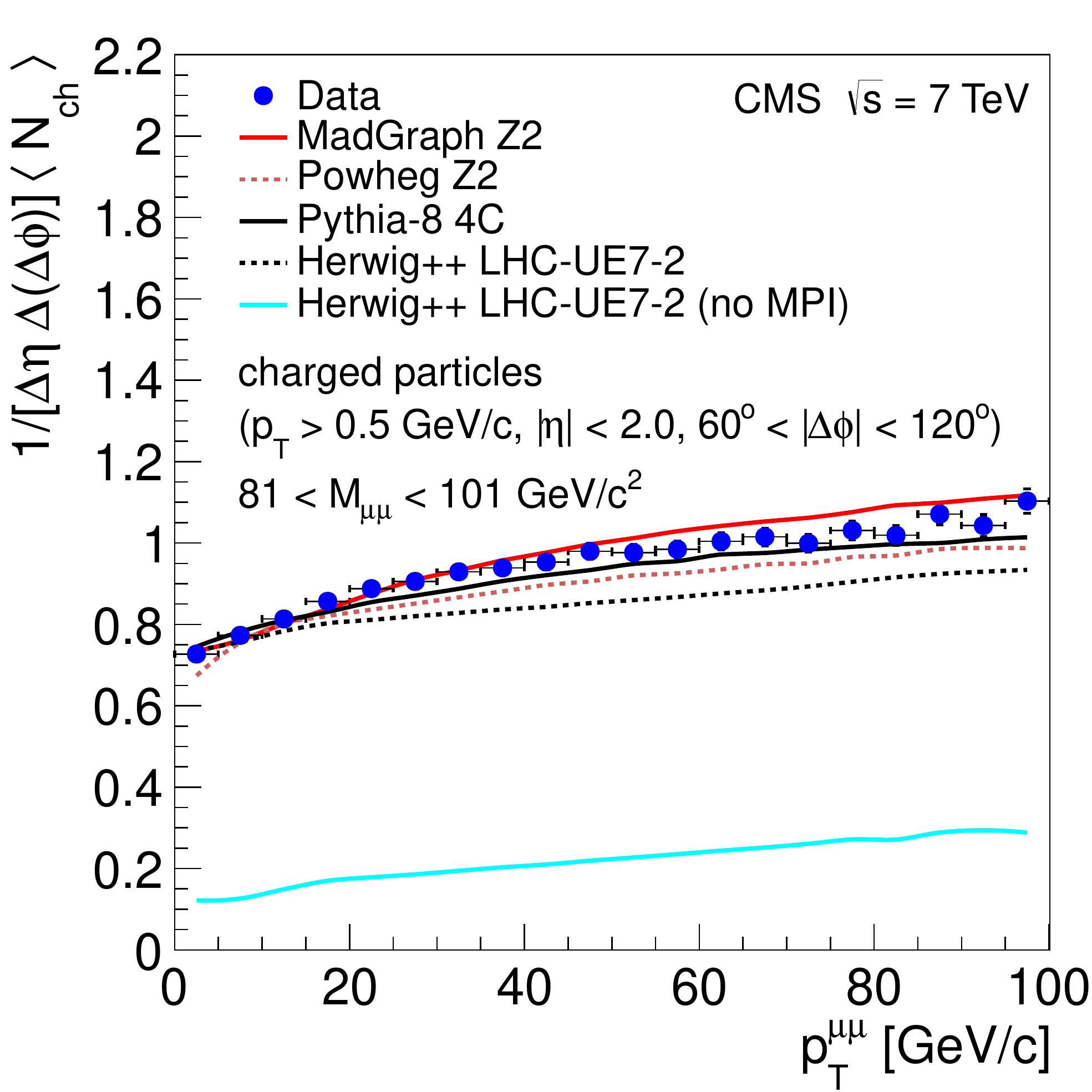}}
\caption{The corrected measurements of charged particles in the transverse region, 60$^\circ$ $<$ $\Delta\phi$ $<$ 120$^\circ$, as a function of the $p_T$ of the leading track for QCD events (a,b) and the dimuon pair for Drell-Yan events (c,d). The average multiplicity per unit of pseudorapidity and azimuthal angle is represented in (a) and (c), while the average scalar per unit of pseudorapidity and azimuthal angle is shown in (b) and (d) for the two measurements. The error bars cover in all of them the statistical and the systematic uncertainties added in quadrature.}
\label{SoftMPIp1}
\end{center}
\end{figure}

In Fig.\ref{SoftMPIp1}, the first two variables are shown for the two different processes. A different behaviour can be noticed: for the QCD events (Fig. \ref{SoftMPIp1}(a) and (b)), the curves, as a function of the leading track $p_T$, show a fast rise up to 4 GeV due to the MPI component and a rather constant slope going to higher $p_T$ where the MPI contribution is saturated and the slight slope is determined by radiation effects. In particular, this behaviour is due to the fact that, with increasing scale of the hard process, the collision occurs more and more central and the probability to have more than one partonic interaction tends to increase: this is valid in the low scale region, until the plateaux is reached, where the MPI contribution remains constant. Drell-Yan processes (Fig. \ref{SoftMPIp1}(c) and (d)), instead, exhibit a continuous slow rise along the whole range of di-muon $p_T$: the MPI contribution is already saturated due to the selected di-muon mass range and the effects of the radiation regulate the slow increase going towards high $p_T$. \\
These two interesting measurements show the hadronic activity that affects two different hard processes: differences have been observed regarding the shapes of the distributions due to the different regimes. The results are based on the detection of single soft tracks and they constitute the starting point to look at a harder UE particle production.

\section{Hard QCD measurements}
A question that is currently addressed in CMS is the possibility for the MPI component to generate hard objects, namely jets, that could be directly accessed by looking at high multiplicity final state. A wide range of physics channels might be useful for this purpose and they are able to scan about 15 orders of magnitude in the process cross sections for the first and second hard scatterings. Theroretical and experimental investigations of the DPS contribution to high energy processes have a long history and are documented in the literature (\cite{Berger:2009cm,Abe:1997bp,Aad:2013bjm}). The whole DPS phenomenology is based on the general expression for the cross section $\sigma^{DPS}_{(A,B)}$: after some assumptions \cite{Ryskin:2011kk}, related to the factorization of the two scatterings, it can be simplified to the following simple quantity:
\begin{equation} 
\sigma^{DPS}_{(A,B)}=\frac{m}{2} \frac{\sigma_A\cdot\sigma_B}{\sigma_{eff}}
\end{equation}
where $\sigma_{eff}$ is a simple proportionality factor that describes the inner structure of the colliding protons. CMS is adopting a three-step strategy in order to perform the extraction of this parameter; first of all, variables sensitive to the DPS contribution, corrected at the stable particle level, are measured. Then, a study of templates for the signal (DPS) and the background (SPS) is performed and the fraction of each of them that best fits the data is obtained. Finally, this fraction is used to extract the $\sigma_{eff}$ parameter. The first step has been accomplished for the W+jets and the 4-jet channels but many other analyses, including $\gamma$+3 jets and 2~b-~+~2~jets, are well in progress: the possibility of a $\sigma_{eff}$ extraction from different channels assures a better understanding of the DPS process dependence. In the following, the measurements of kinematical observables sensitive to DPS are described for the W+jets and the 4 jet channel.\\  
A DPS process leads to a W~+~2~jets final state when one hard interaction produces a W boson and the second one produces 2 jets in the same collisions. Events containing W~+~2~jets produced from a single parton scattering (SPS) contribute an irreducible background. This analysis makes use of the full 2011 data sample, corresponding to an integrated luminosity of 5 $fb^{-1}$ and requires the presence of at least one muon candidate at the trigger level. The selection criteria are summarized in table \ref{tab:sel1}. 

\begin{table}[htbp]
\caption{\label{tab:sel1} Phase space definition for visible cross section at particle-level. }
\begin{center}
\begin{tabular}{c c} \hline
Exactly one muon with & $\ensuremath{\not\!\!E_T} > 30 $ GeV/c\\\ 
$p_{T} > 35$ GeV/c and $|\eta| < 2.1$ & \\
\hline
W transverse mass $ > 50$ GeV/c & jets with $p_{T} > 20$ GeV/c, $|\eta| < 2.0$   \\
\hline
\end{tabular}
\end{center}
\end{table}

The W candidate events are required to have exactly one muon with $p_T$ $>$ 35 GeV/c and $|\eta|$ $<$ 2.1. These events are further required to have $\ensuremath{\not\!\!E_T} >$ 30 GeV and $M_T$ $>$ 50 GeV. The two selected jets are clustered with the anti-$k_T$ jet algorithm \cite{Cacciari:2005hq,Cacciari:2008gp,Cacciari:2011ma} with a distance parameter of 0.5, and are required to have $|\eta|$ $<$ 2.0 and $p_T$ $>$ 20 GeV/c.
Several observables which might be sensitive to discriminate DPS events from the SPS ones have been defined:
\begin{itemize}
\item the azimuthal separation between the two selected jets: \\
\vspace{-8mm}\begin{center} $\Delta\phi$ = $\phi_{j1}$-$\phi_{j2}$ \end{center}
\item the normalized $p_T$ balance between the two selected jets:\\ \vspace{-8mm} \begin{center} $\Delta^{rel} p_T$ = $\frac{|\vec{p_T}(j1)+\vec{p_T}(j2)|}{|\vec{p_T}(j1)|+|\vec{p_T}(j2)|}$ \end{center}
\item the azimuthal angle between the W and the dijet system: \\ \vspace{-8mm} \begin{center} $\Delta S$=$\arccos\left(\frac{\vec{P_T}(\mu, \ensuremath{\not\!\!E_T}\cdot \vec{P_T}(j1,j2)}{|\vec{P_T}(\mu, \ensuremath{\not\!\!E_T})|\cdot |\vec{P_T}(j1,j2)|}\right)$ \end{center}
\end{itemize}

\begin{figure}[htbp]
\begin{center}
\subfigure[]{\includegraphics[width=4cm]{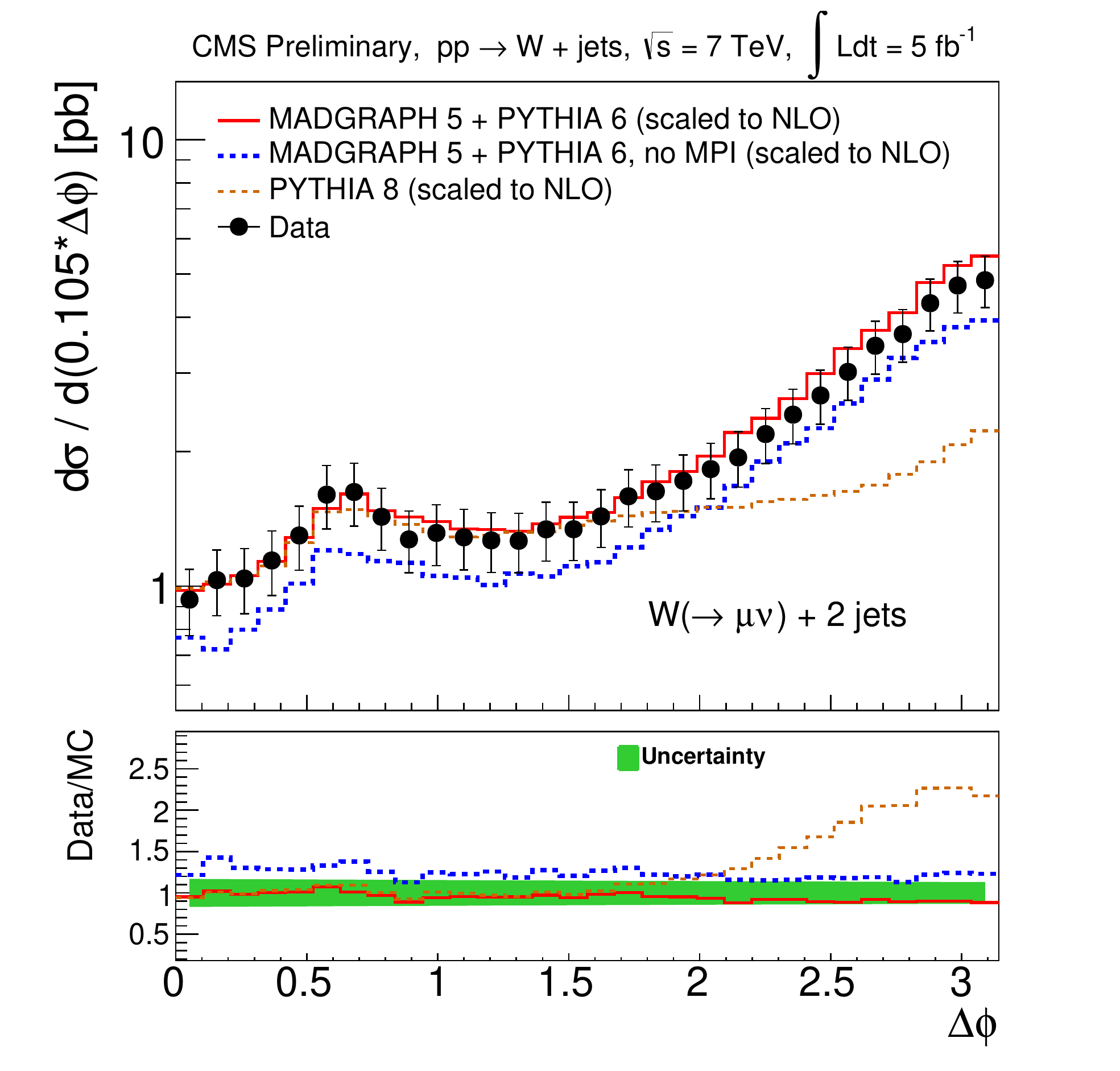}}
\subfigure[]{\includegraphics[width=4cm]{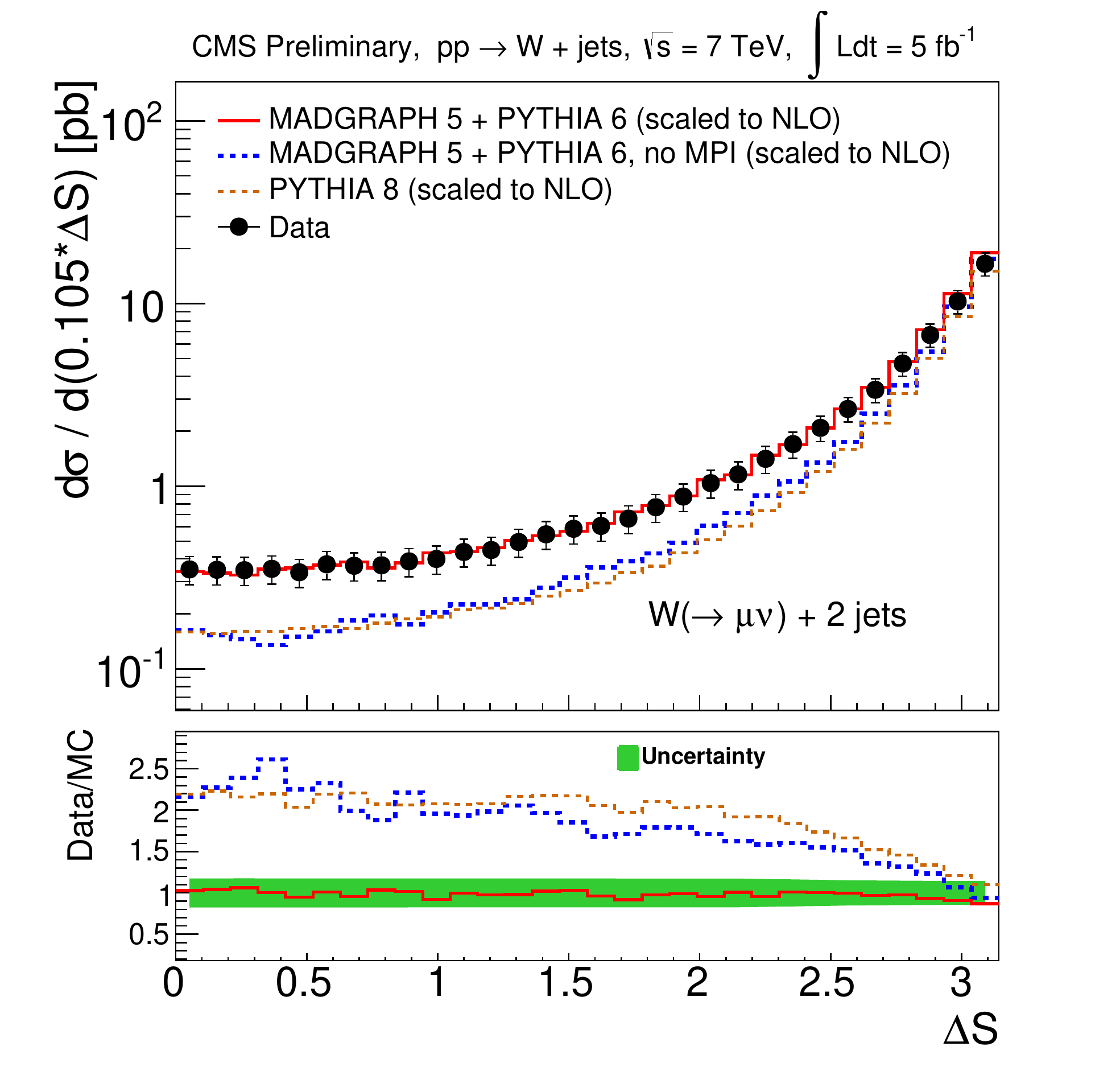}}
\subfigure[]{\includegraphics[width=4cm]{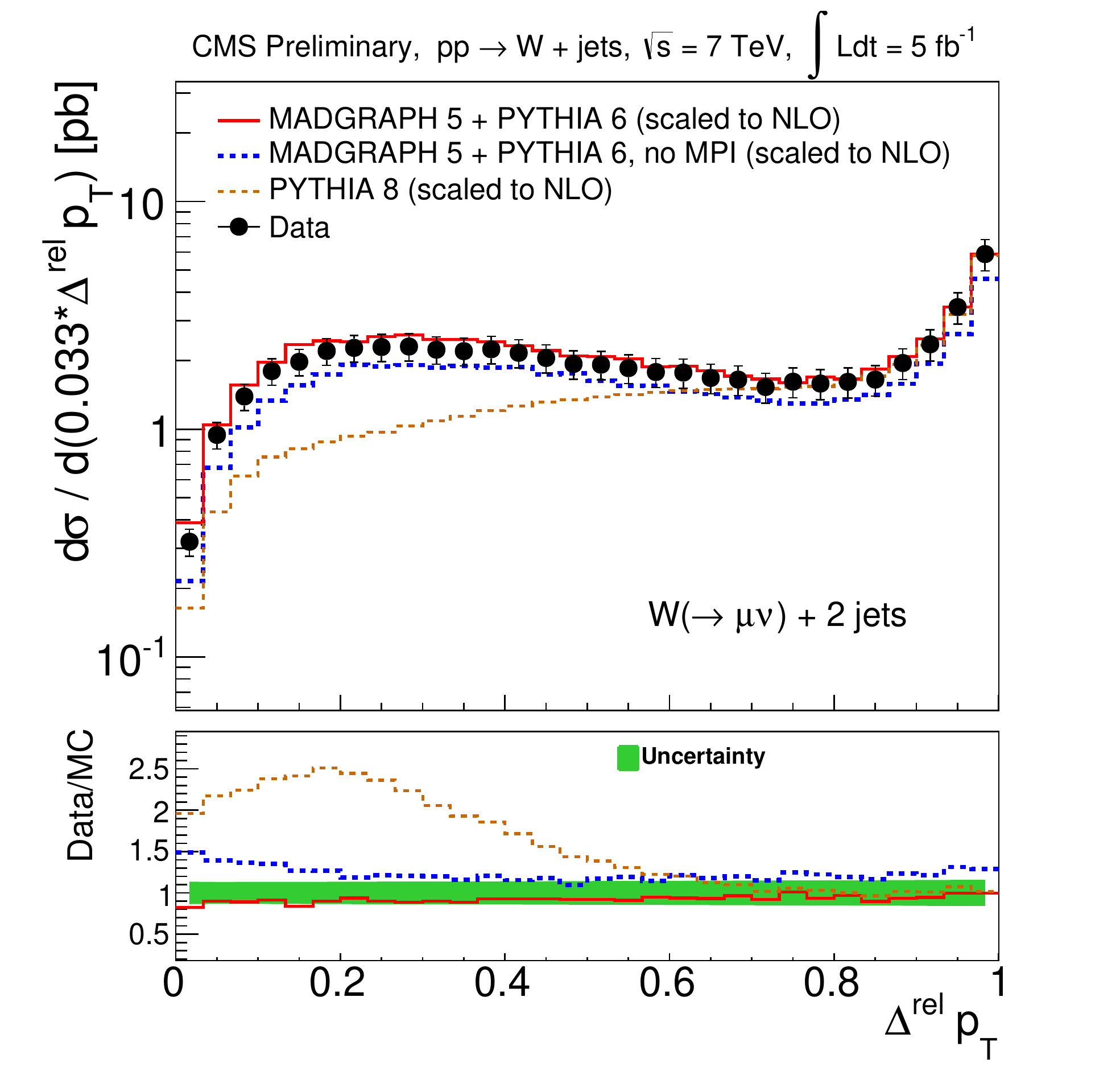}}\\
\subfigure[]{\includegraphics[width=4cm]{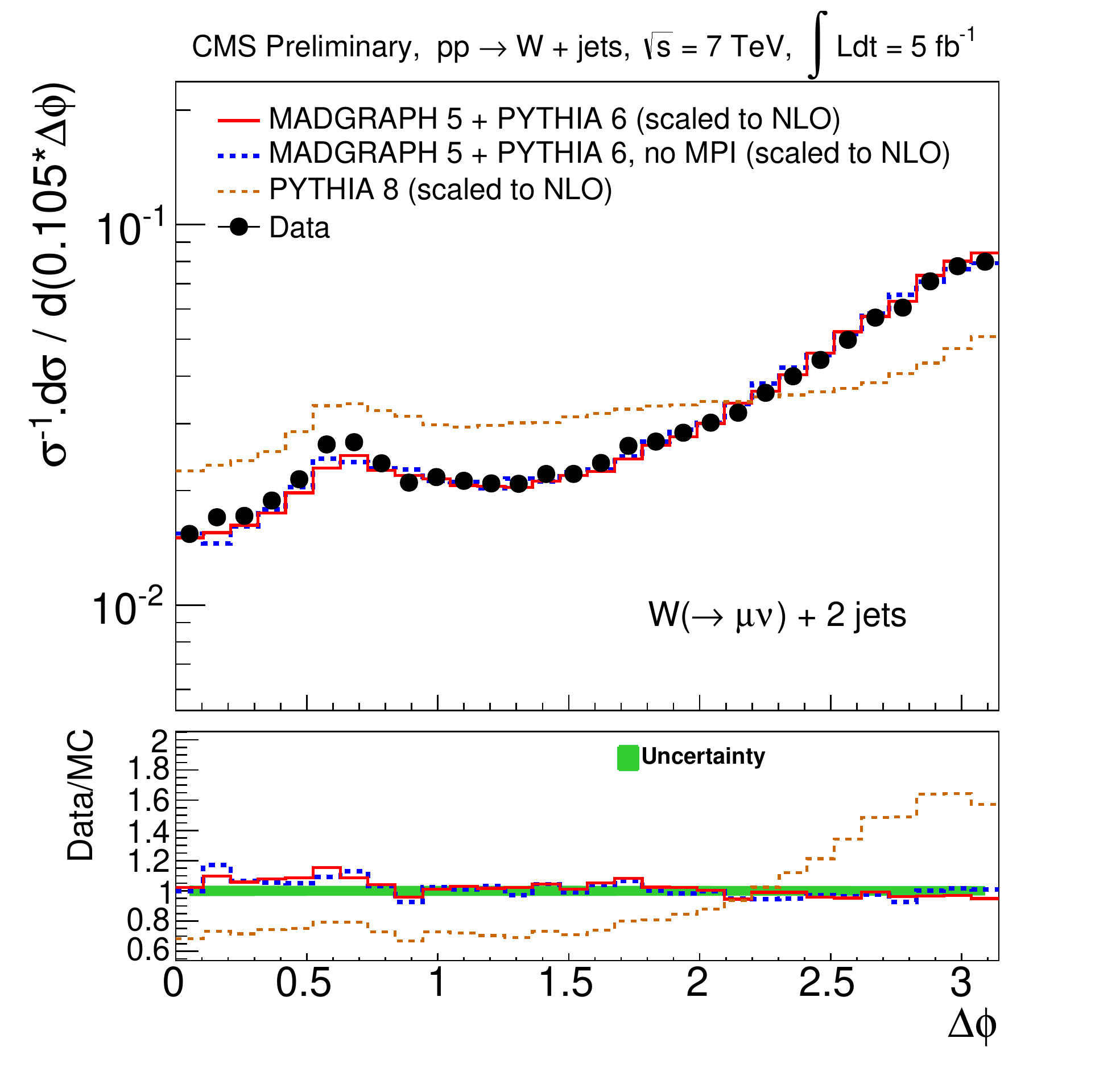}}
\subfigure[]{\includegraphics[width=4cm]{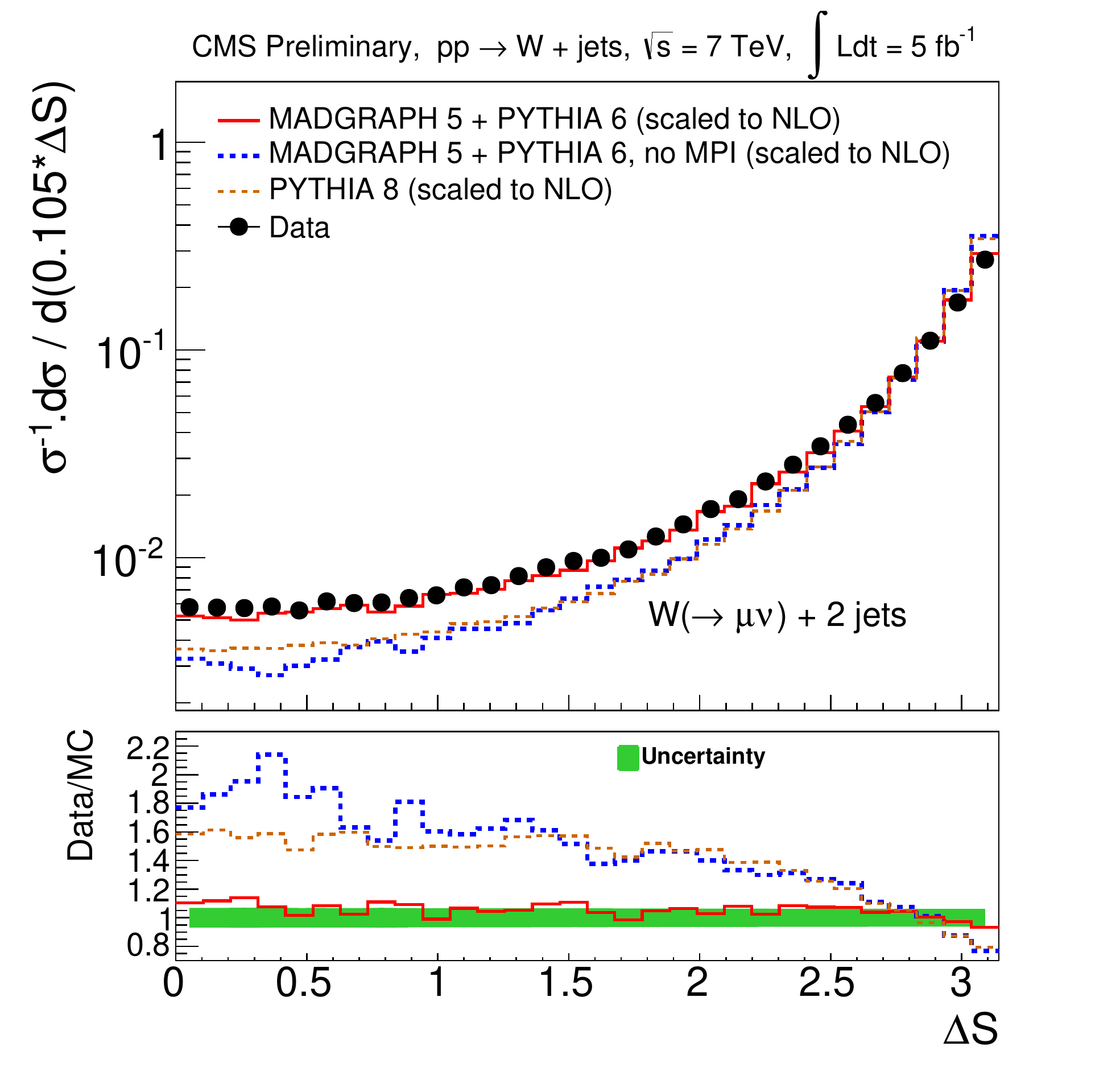}}
\subfigure[]{\includegraphics[width=4cm]{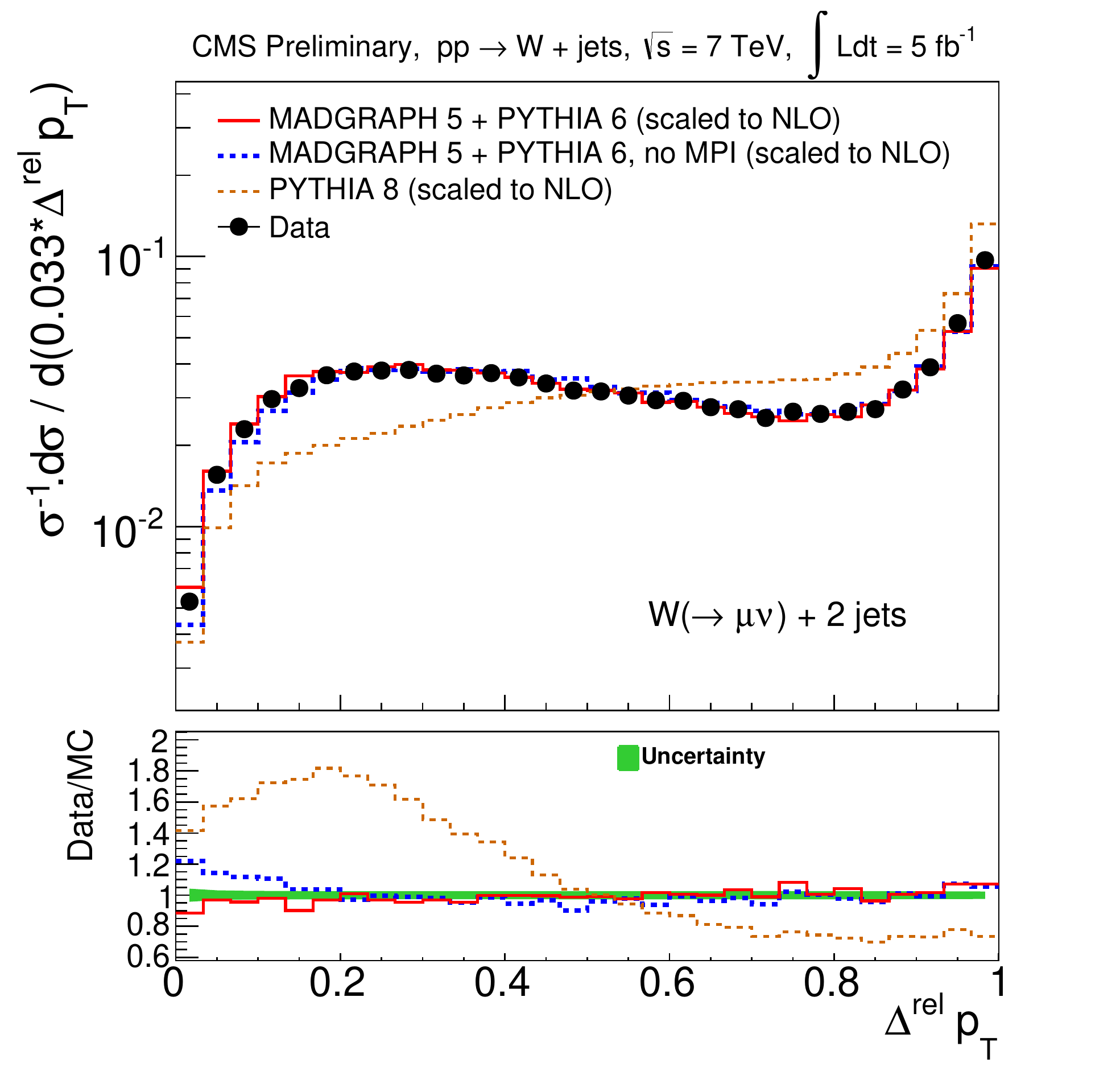}}\\
\caption{Fully corrected differential cross sections and normalized distributions for various DPS-sensitive observables: $\Delta^{rel}p_T$ (a,d), $\Delta\phi$ (b,e), and $\Delta S$ (c,f). The bottom pad in each plot shows the ratio of data over simulations. The green band represents the total uncertainty in the experimental distribution.}
\label{FSQ12028}
\end{center}
\end{figure}

Figure \ref{FSQ12028} shows the differential cross section and normalized distributions of DPS-sensitive observables after applying full correction for detector effects and selection efficiency. Corrected distributions are compared to particle-level predictions of the \textsc{Madgraph} MC generator. This sample includes W+jets (up to 4 partons) hard processes from the matrix element calculation which are showered and hadronized using \textsc{Pythia 6} Z2* \cite{field2010} tune. DPS-sensitive distributions are nicely described by \textsc{Madgraph} predictions. The Monte Carlo predictions without MPI do not describe the measurement. The effect of MPI is not only visible in the cross section values: also the shape of the distributions is not in agreement with measurements if MPI is not included. The different observables have different sensitivity in separating DPS contributions from SPS. The shape of the $\Delta\phi$ distribution is almost unaffected when MPI is turned off whereas $\Delta S$ shows a deviation of 40-100\% in the DPS-sensitive region. \textsc{Pythia 8} underestimates the measurements by a factor 1.5-2, especially in the DPS-sensitive region. This discrepancy is not due to wrong or missing contribution of DPS, but this is the effect of missing hard process contribution as \textsc{Pythia 8} generate only 2$\rightarrow$1 and 2$\rightarrow$2 process.
 Thus, there is a significant effect of higher-order hard processes in DPS-sensitive regions. It is important that the samples used to define the signal and background templates, while extracting DPS fractions, have proper inclusion of additional hard partons. Otherwise missing hard process contributions might be interpreted as DPS. These variables have been measured also for two further inclusive scenarios where more than two jets are accepted: in particular, in one case, the two leading jets are associated and in the other one, the two best-balanced jets are considered \cite{CMS:awa}. \\

Another important channel to search for a DPS contribution is the exclusive final state with four light jets, associated in two pairs with different $p_T$. In this case, the two pair of jets can be produced either by a SPS with two hard jets emitted by the first scattering and two soft jets generated through parton shower or by a DPS, where the two different chains may produce the dijet systems. In order to separate the two contributions, kinematical variables related to the different jet pairs, as done for the W+jets channel are measured. Furthermore, to set different thresholds for the jets in the final state gives the opportunity to test the predictions of different types of MC generators, like pure leading-order (\textsc{Pythia} and \textsc{Herwig}), next-to-leading order (\textsc{Powheg}) and higher-order matrix element (\textsc{Madgraph}, \textsc{Sherpa}) ones. An integrated luminosity of 36 pb$^{-1}$ from 2010 data has been used and single jet triggers have been applied. A final state with exactly four jets in the final state has been selected: for two of them, a $p_T$ threshold of 50 GeV has been set: they form the ``hard-jet'' pair. The other two, the ``soft-jet'' pair, are required to have $p_T$ greater than 20 GeV. All the jets are required to be in the central region, defined as $|\eta|$ $<$ 2.5, and the jet clustering is based on the anti-$k_T$ jet algorithm with R = 0.5 \cite{Cacciari:2005hq,Cacciari:2008gp,Cacciari:2011ma}. 
       
\begin{figure}[htbp]
\begin{center}
\subfigure[]{\includegraphics[width=3.29cm]{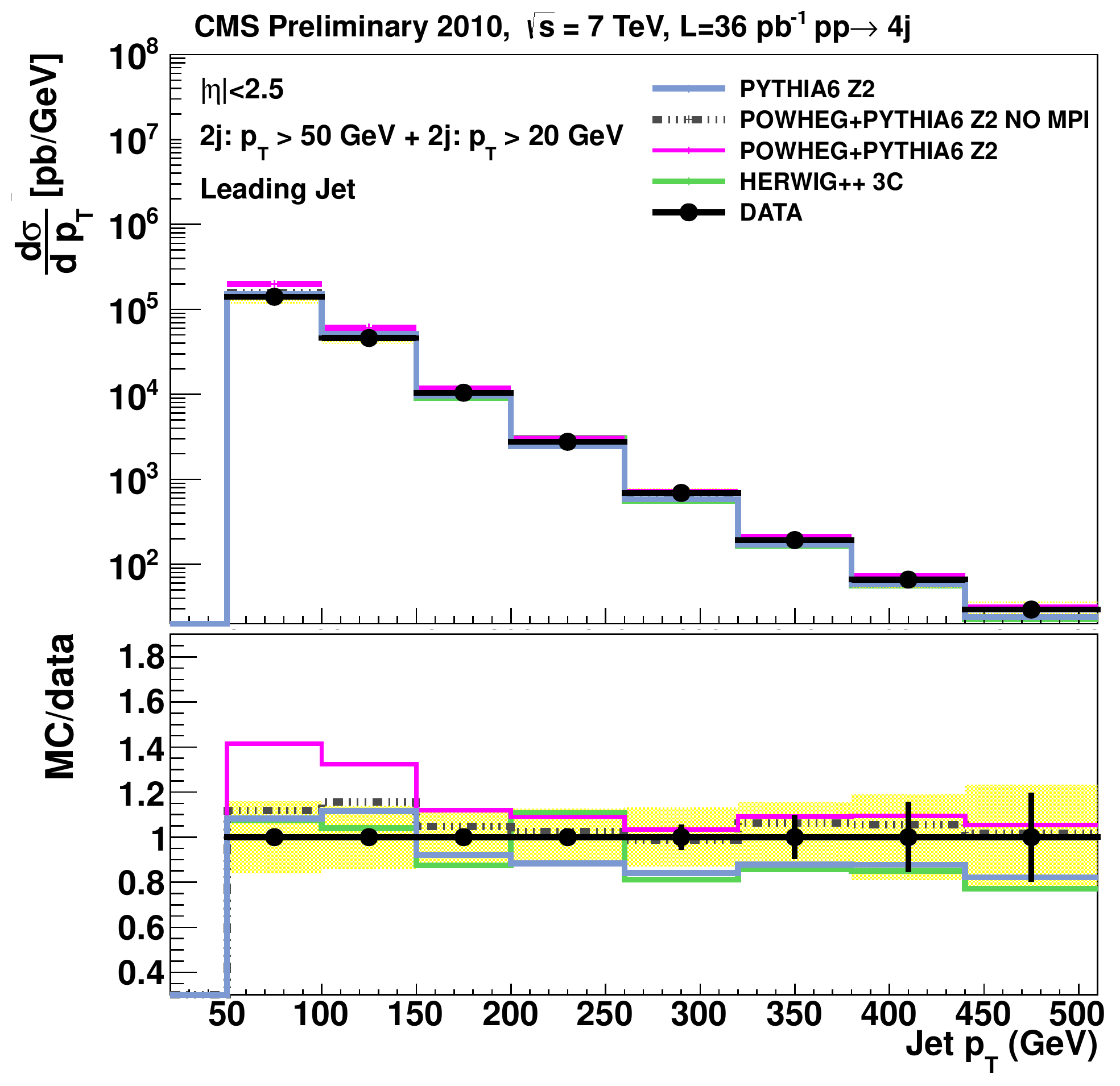}}
\subfigure[]{\includegraphics[width=3.29cm]{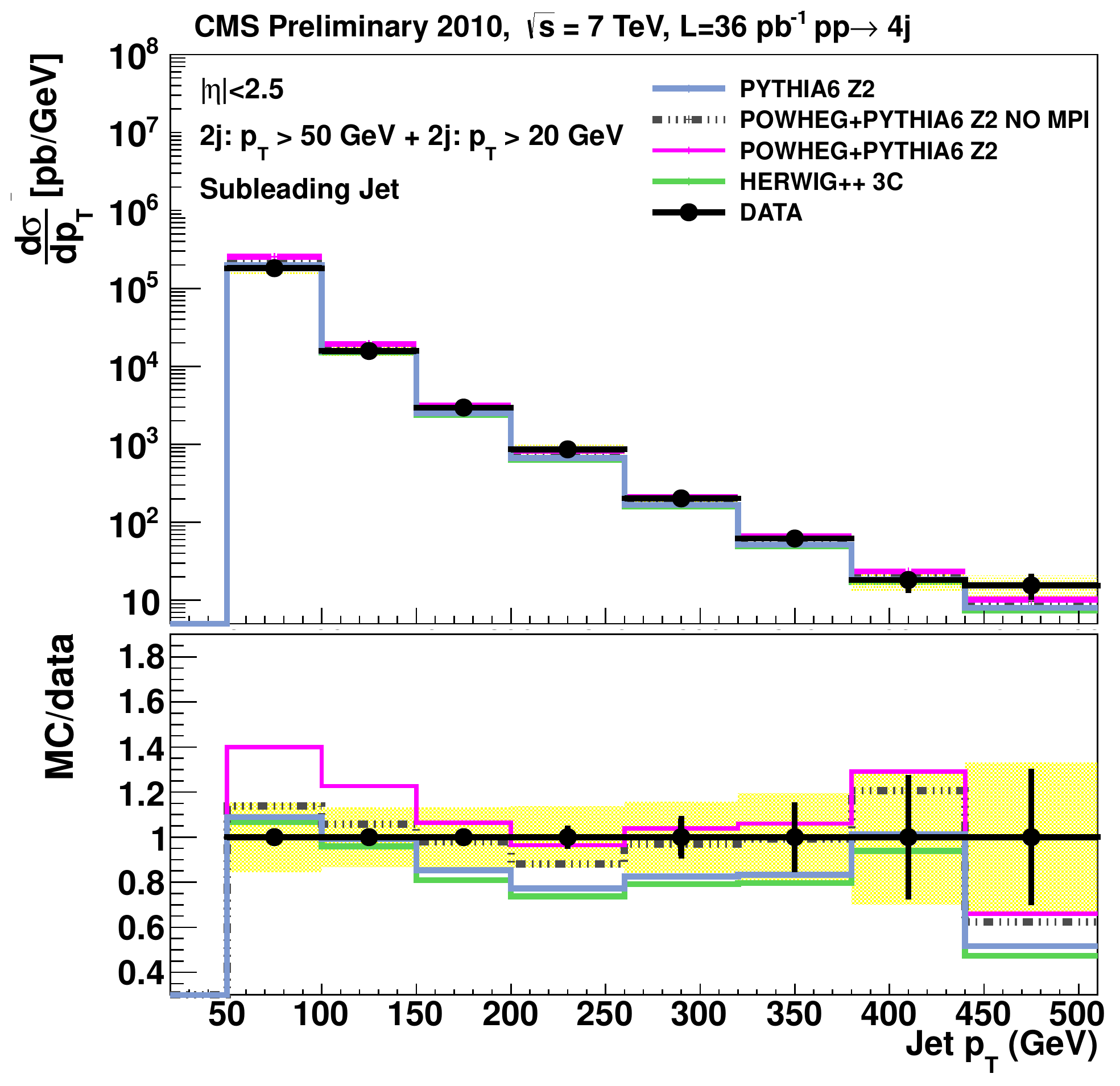}}
\subfigure[]{\includegraphics[width=3.29cm]{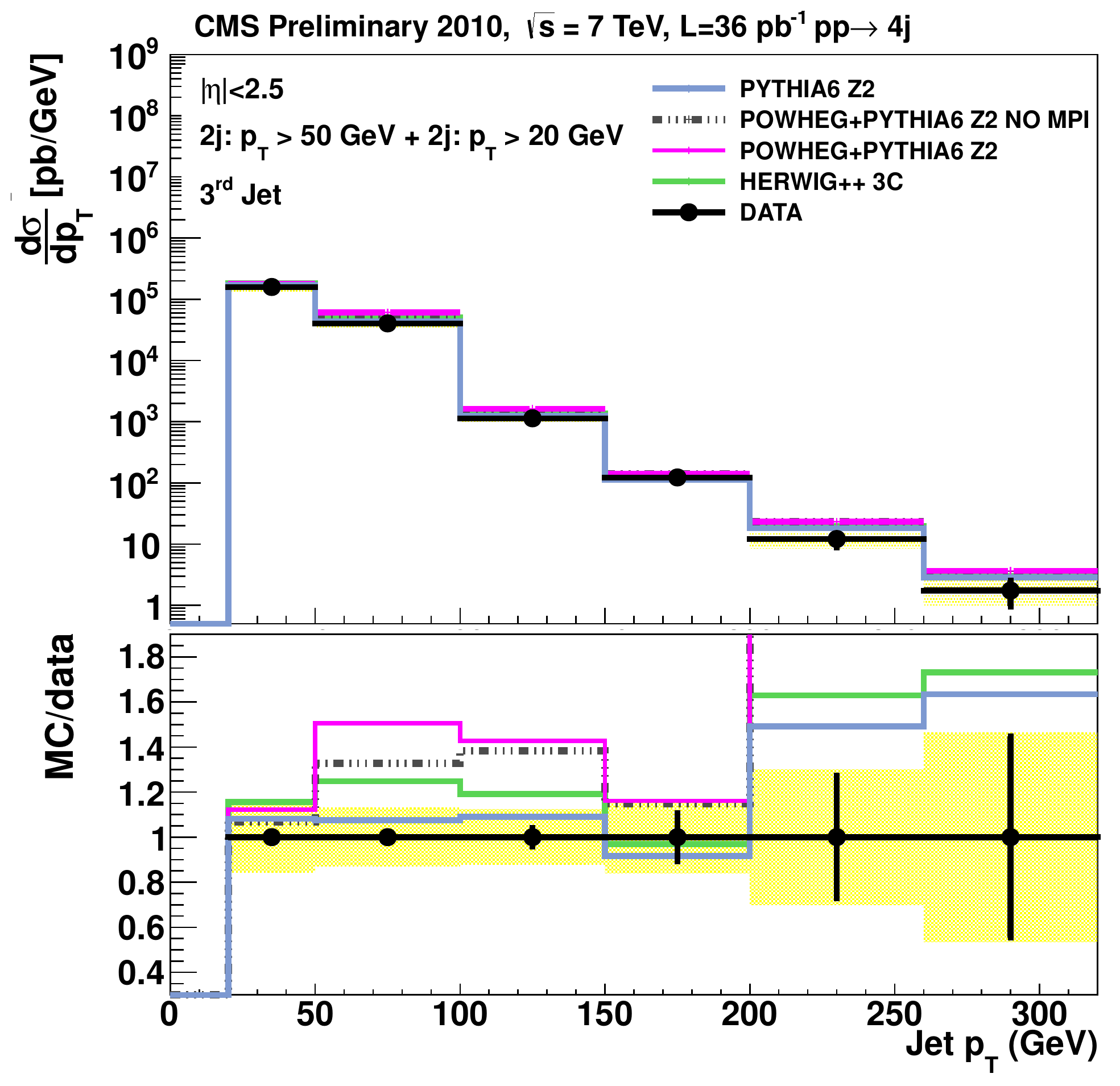}}
\subfigure[]{\includegraphics[width=3.29cm]{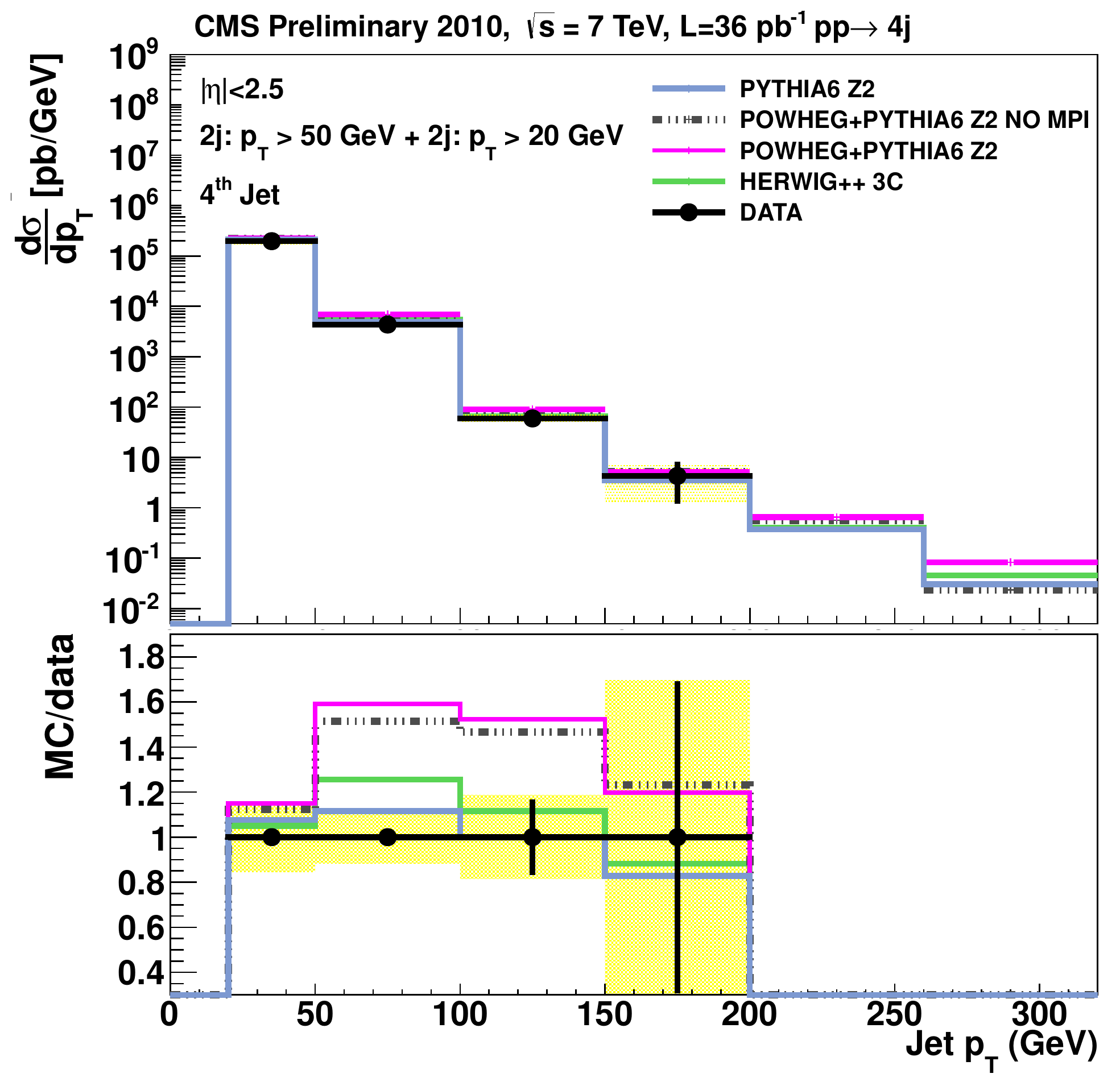}}\\
\caption{Cross section as a function of the jet transverse momenta $p_T$ (from (a) to (d) leading, subleading, third and forth jet) compared to theoretical predictions of \textsc{Pythia 6}, \textsc{Herwig++} and \textsc{Powheg}. \textsc{Powheg} is shown with and without contributions from MPI.  The lower panel shows the ratio of theory prediction to data.}
\label{FSQ12013p1}
\end{center}
\end{figure}

The cross sections are fully corrected for detector effects and efficiencies. The cross sections measured as a function of the transverse momenta $p_T$ of the four jets are shown in Fig. \ref{FSQ12013p1}. The cross sections are falling rapidly with increasing $p_T$ for all the jets in the final state: for the highest $p_T$ jets (Fig. \ref{FSQ12013p1}(a) and (b)) it decreases over two orders of magnitude for $p_T$ between 50 and 200 GeV, while the cross section for the softer jets decreases over 5 orders of magnitude in the same $p_T$ range. The spectrum of the 3rd jet at low transverse momenta (20-80 GeV) is rather flat while falling rapidly for $p_T$ $>$ 100 GeV (Fig. \ref{FSQ12013p1}(c)), a consequence of the asymmetric $p_T$ thresholds for the two jet pairs.

\begin{figure}[htbp]
\begin{center}
\subfigure[]{\includegraphics[width=3.29cm]{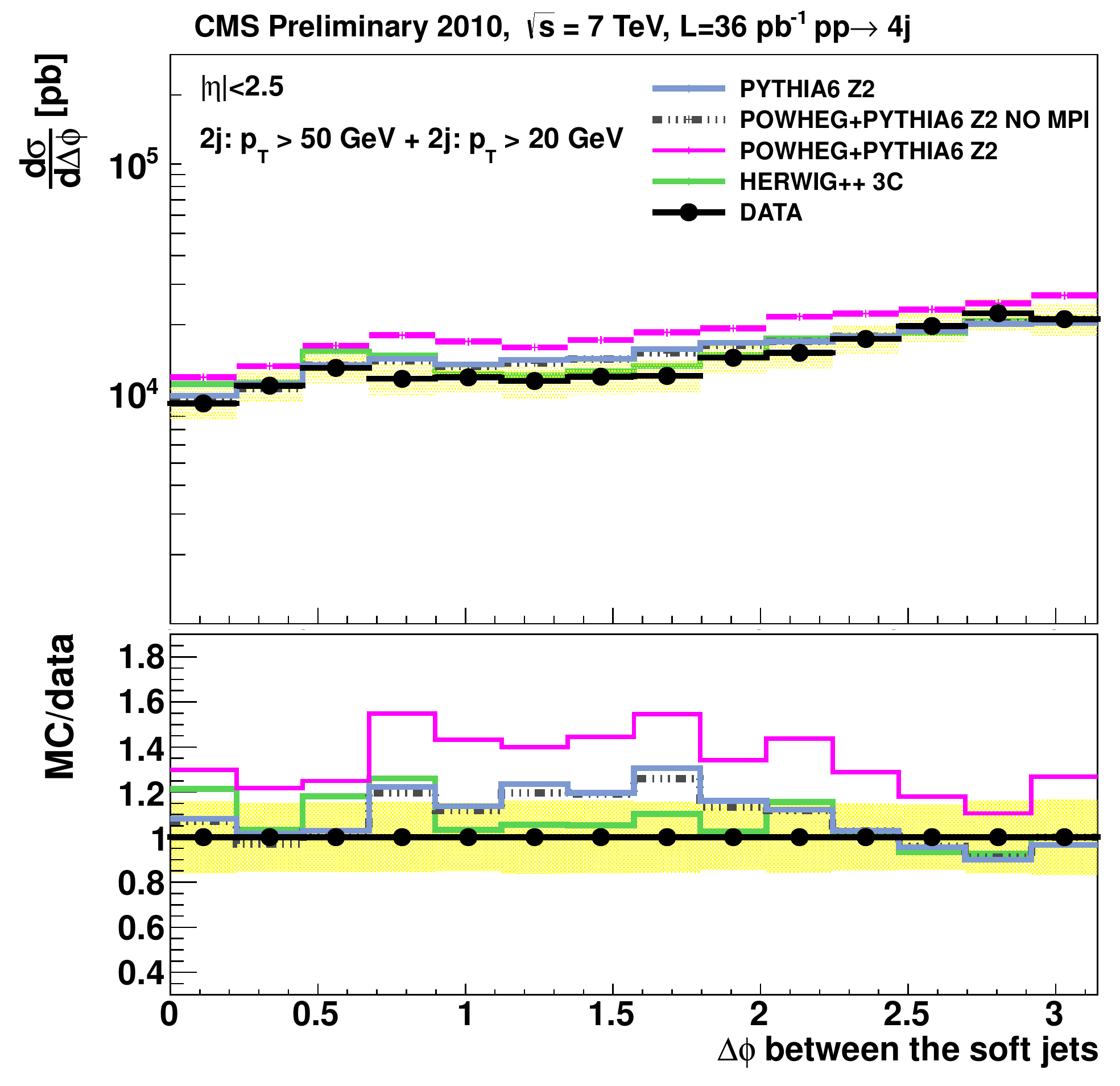}}
\subfigure[]{\includegraphics[width=3.29cm]{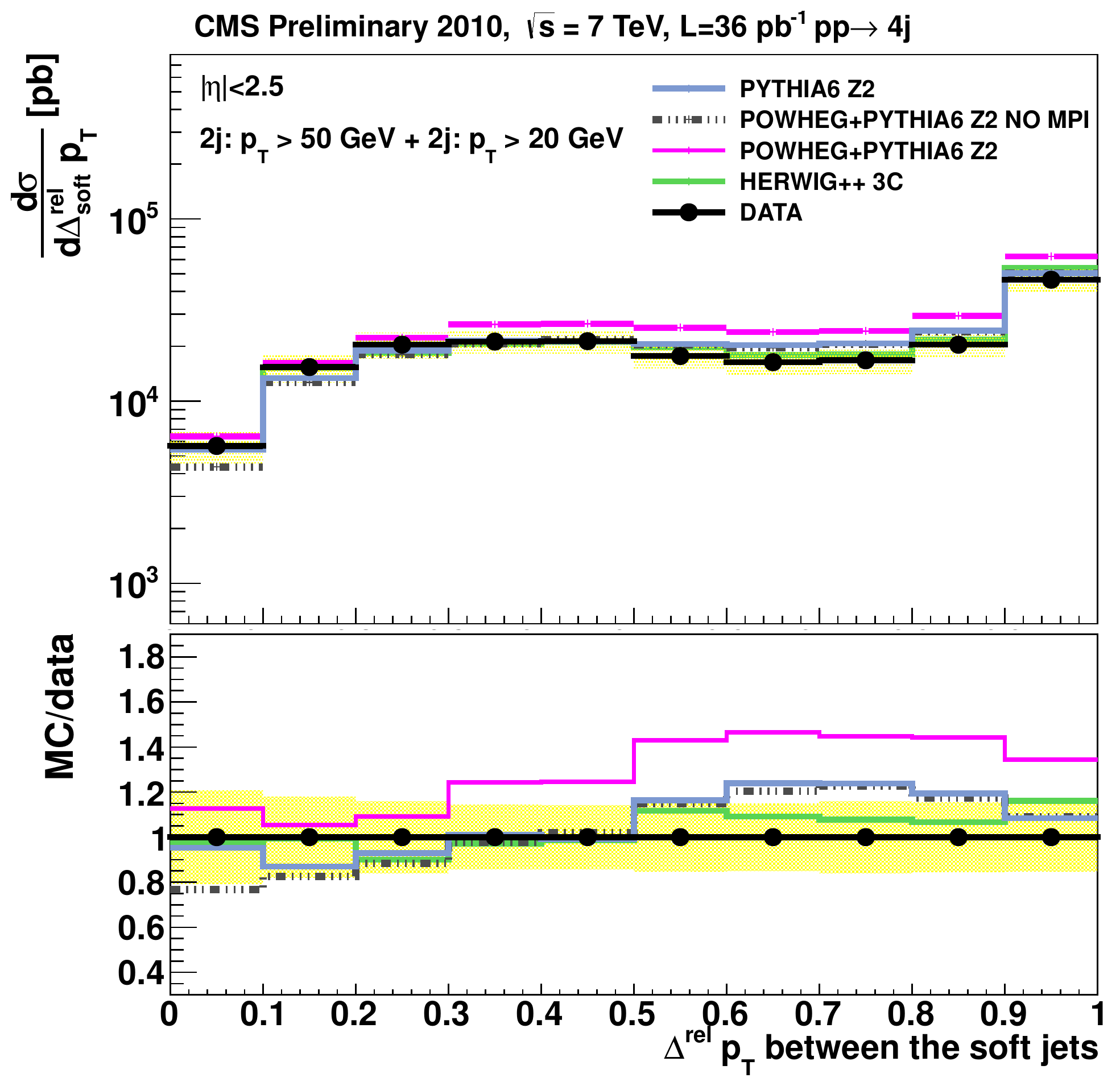}}
\subfigure[]{\includegraphics[width=3.29cm]{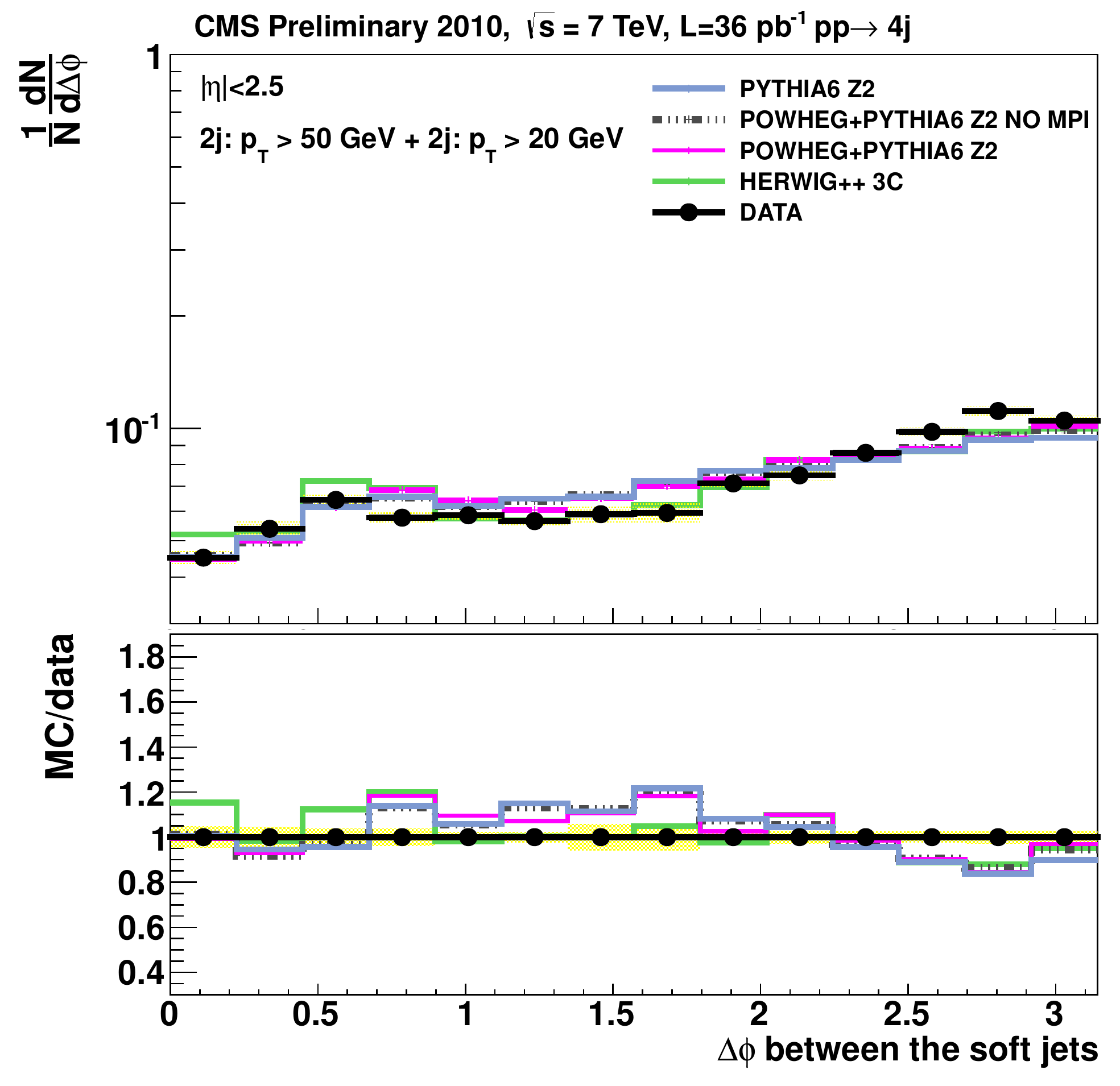}}
\subfigure[]{\includegraphics[width=3.29cm]{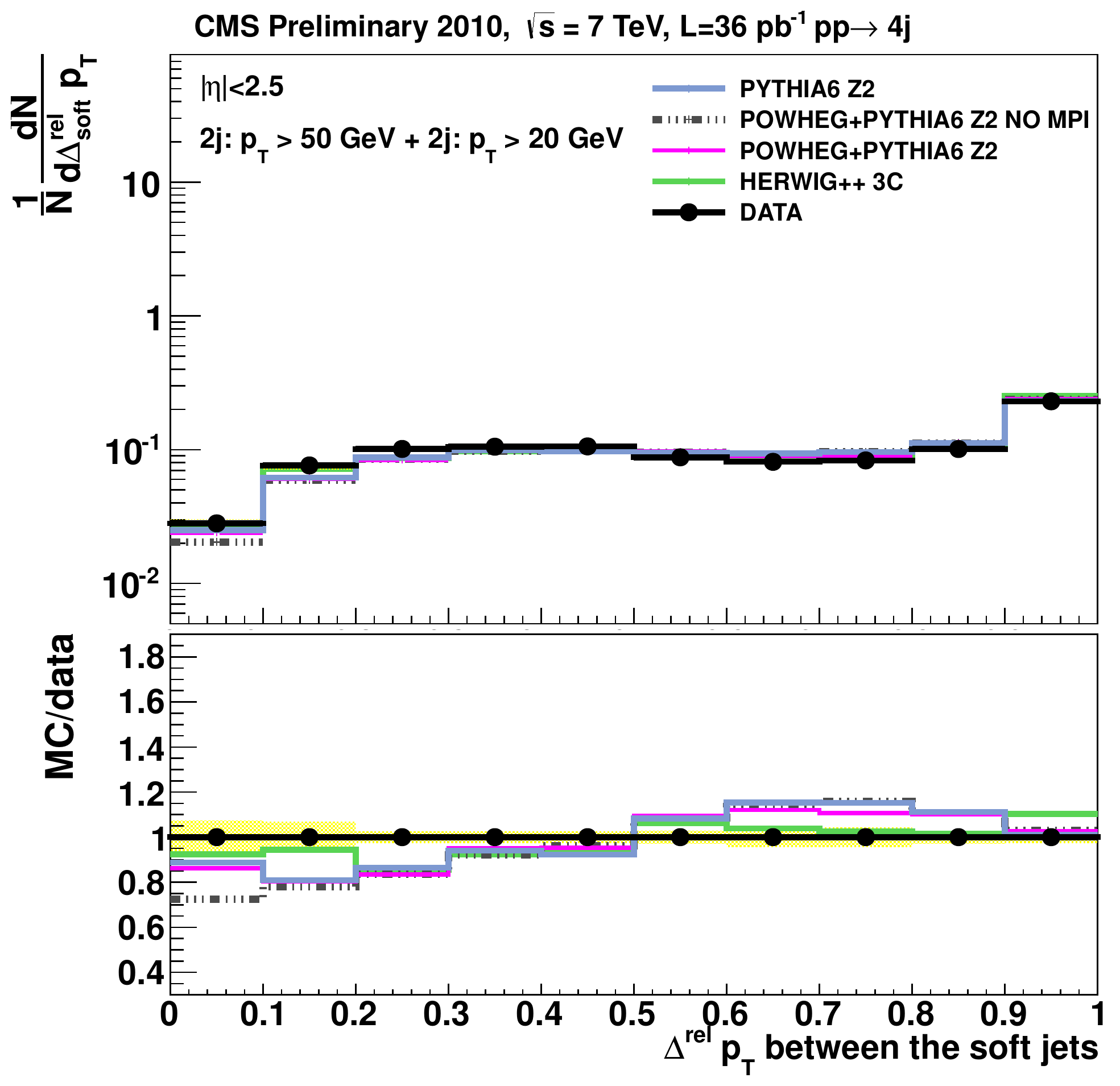}}\\
\caption{Cross section (a,b) and normalized distributions (c,d) as a function of $\Delta\phi_{soft}$ (a,c) and of $\Delta^{rel}_{soft}p_T$ (b,d) compared to theoretical predictions of \textsc{Pythia 6}, \textsc{Herwig ++} and \textsc{Powheg}. \textsc{Powheg} is shown with and without contributions from MPI.  The lower panel shows the ratio of theory prediction to data. The yellow band represents the total uncertainty, while the black line shows the statistical error.}
\label{FSQ12013p2}
\end{center}
\end{figure}

The comparison between data and Monte Carlo predictions shows an overall good agreement in the measured observables. In general, \textsc{Powheg} interfaced with \textsc{Pythia}~Z2 tends to overshoot the data in the low $p_T$ region (up to 150 GeV) of the hard jets (Fig. \ref{FSQ12013p1}(a,b)) and the whole $p_T$ range of the soft jets (Fig. \ref{FSQ12013p1}(c,d)). The predicted excess is around 30-50\%. The agreement between measurement and \textsc{Powheg} with \textsc{Pythia 6} Z2 as a function of the leading jet $p_T$ improves significantly if the contribution of the MPI is switched off. \\This is a clear indication that the best tune for \textsc{Pythia 6} does not perform in the same good way when interfaced with a NLO MC generator, like \textsc{Powheg}.
The cross sections and the normalized distributions as a function of $\Delta\phi$ and $\Delta^{rel}p_T$ for the soft jet pair are shown in Fig. \ref{FSQ12013p2}. The $\Delta\phi$ distribution has a maximum at about $\pi$ and falls over an order of magnitude towards very small values. At small $\Delta\phi$ the jets are de-correlated. The data are reasonably well described by the theoretical predictions. Regarding the $p_T$ balance, the soft jets have a maximum at around 1, indicating that they are predominantly not balanced in transverse momentum , which is the case if they come from radiation of the initial or final state of the hard jets. However the distribution exhibits another maximum around 0.2. Interesting is to observe, that the shapes for both $\Delta\phi$ and $\Delta^{rel}p_T$ are reasonably well reproduced even by \textsc{Powheg} interfaced with \textsc{Pythia 6} Z2. The full set of results is reported in \cite{CMS:mine}.\\
These two analyses have set an important baseline for the extraction of the DPS contribution but they have also shown the necessity of a better understanding of the SPS process where the effects of higher-order matrix elements in the correlation observables are not yet completely understood, specifically in the regions of the phase space where the DPS may play a role.

\section{Summary}
A brief overview of the ongoing CMS research on multiparton interactions has been presented. This basically deals with two categories of events, related to the scale of the additional partonic interactions that occur : soft and semi-hard MPI can be studied through the detection of the hadronic activity in the transverse region with respect to the hard scattering, while the presence of additional jets in particular regions of the phase space might be an indication of the hard MPI contribution. The measurement of the underlying event component in QCD and Drell-Yan events showed some deviations, related to the different shape of the observables in the low-$p_T$ region. The measurement of kinematical correlations in the W+jets and in the 4 light-jet channels has been carried out in order to study the possibility to have a hard jet production from MPI and provides a useful baseline for future studies to investigate possible contributions from DPS.

\bibliography{ProceedingGunnellini}

\begin{thebibliography}{10}

\bibitem{Bartalini:2011jp}
P.~Bartalini, E.L. Berger, B.~Blok, G.~Calucci, R.~Corke, et~al.
\newblock {Multi-Parton Interactions at the LHC}.
\newblock {\em arXiv}, 1111.0469, 2011.

\bibitem{Chatrchyan:2008aa}
S.~Chatrchyan et~al.
\newblock {The CMS experiment at the CERN LHC}.
\newblock {\em JINST}, 3:S08004, 2008.

\bibitem{CMS:2012zxa}
CMS Collaboration.
\newblock {Measurement of the Underlying Event Activity at the LHC at 7 TeV and
  Comparison with 0.9 TeV}.
\newblock {\em CMS-PAS-FSQ-12-020}, 2012.

\bibitem{Chatrchyan:2012tb}
Serguei Chatrchyan et~al.
\newblock {Measurement of the underlying event in the Drell-Yan process in
  proton-proton collisions at $\sqrt{s}=7$ TeV}.
\newblock {\em Eur.Phys.J.}, C72:2080, 2012.

\bibitem{ALICE:2011ac}
Betty Abelev et~al.
\newblock {Underlying Event measurements in $pp$ collisions at $\sqrt{s}=0.9$
  and 7 TeV with the ALICE experiment at the LHC}.
\newblock {\em JHEP}, 1207:116, 2012.

\bibitem{Berger:2009cm}
Edmond L. et~al. Berger.
\newblock {Characteristics and Estimates of Double Parton Scattering at the
  Large Hadron Collider}.
\newblock {\em Phys.Rev.}, D81:014014, 2010.

\bibitem{Abe:1997bp}
F.~Abe et~al.
\newblock {Measurement of double parton scattering in $\bar{p}p$ collisions at
  $\sqrt{s} = 1.8$ TeV}.
\newblock {\em Phys.Rev.Lett.}, 79:584--589, 1997.

\bibitem{Aad:2013bjm}
Georges Aad et~al.
\newblock {Measurement of hard double-parton interactions in
  W$\rightarrow$l$\nu$ + 2 jet events at sqrt(s)=7 TeV with the ATLAS
  detector}.
\newblock {\em New J.Phys.}, 15:033038, 2013.

\bibitem{Ryskin:2011kk}
M.G. Ryskin and A.M. Snigirev.
\newblock {A Fresh look at double parton scattering}.
\newblock {\em Phys.Rev.}, D83:114047, 2011.

\bibitem{Cacciari:2005hq}
Matteo Cacciari and Gavin~P. Salam.
\newblock {Dispelling the $N^{3}$ myth for the $k_t$ jet-finder}.
\newblock {\em Phys.Lett.}, B641:57--61, 2006.

\bibitem{Cacciari:2008gp}
Matteo Cacciari, Gavin~P. Salam, and Gregory Soyez.
\newblock The anti-$k_t$ jet clustering algorithm.
\newblock {\em JHEP}, 04:063, 2008.

\bibitem{Cacciari:2011ma}
Matteo Cacciari, Gavin~P. Salam, and Gregory Soyez.
\newblock {FastJet User Manual}.
\newblock {\em Eur.Phys.J.}, C72:1896, 2012.

\bibitem{field2010}
R.~Field.
\newblock Early lhc underlying event data - findings and surprises.
\newblock {\em arXiv}, 1010.3558, 2010.

\bibitem{CMS:awa}
CMS Collaboration.
\newblock {Study of observables sensitive to double parton scattering in W + 2
  jets process in p-p collisions at sqrt(s) = 7 TeV}.
\newblock {\em CMS-PAS-FSQ-12-028}.

\bibitem{CMS:mine}
CMS Collaboration.
\newblock {Studies of 4-jet production in proton-proton collisions at $\sqrt{s}
  = 7 $TeV}.
\newblock {\em CMS-PAS-FSQ-12-013}.

\end{thebibliography}

%

 

\end{document}